\documentclass[showpacs,aps,prb,twocolumn,groupedaddress]{revtex4}

\usepackage{graphicx}% Include figure files

\begin{document}

%\preprint{}

\title{Mott transition and superconductivity in the strongly correlated organic superconductor $\kappa$-(BEDT-TTF)$_{2}$Cu[N(CN)$_{2}]$Br}


\author{T. Sasaki}
\email{takahiko@imr.tohoku.ac.jp}
\author{N. Yoneyama}
\author{N. Kobayashi}
%\thanks{}
\affiliation{Institute for Materials Research, Tohoku University, Katahira 2-1-1, Aoba-ku, Sendai 980-8577, Japan}


\date{\today}

\begin{abstract}
The magnetic field effect on the phase diagram of the organic Mott system $\kappa$-(BEDT-TTF)$_{2}$Cu[N(CN)$_{2}$]Br in which the bandwidth was tuned by the substitution of deuterated molecules was studied by means of the resistivity measurements performed in magnetic fields.  
The lower critical point of the first-order Mott transition, which ended on the upper critical field $H_{\rm c2}$-temperature plane of the superconductivity, was determined experimentally in addition to the previously observed upper critical end point.  
The lower critical end point moved to a lower temperature with the suppression of $T_{\rm c}$ in magnetic fields and the Mott transition recognized so far as the $S$-shaped curve reached $T =$ 0 when $H > H_{\rm c2}$ in the end.  

\end{abstract}

%\keywords{}

\maketitle


%Introduction
Metal-insulator (MI) transitions have been one of the central issues in strongly correlated electron systems.
Among the various types of MI transitions, the Mott transition is the most attractive phenomenon, which arises from an electron-electron interaction in a wide range of materials.\cite{Imada}
Mott insulator derives from a large electrical Coulomb repulsion energy $U$ with respect to the bandwidth $W$. 
Metal transitions can occur in two ways in a Mott insulator: one can change the strength of the interaction $U/W$, while maintaining the required band-filling value and is referred to as a bandwidth-controlled Mott transition or one can introduce carriers with the required density, while maintaining $U/W$ and is referred to as a band-filling-controlled Mott transition.  
The former case occurs, for example, in vanadium oxides and molecular conductors, while the latter occurs in high-$T_{\rm c}$ cuprate superconductors.  

Organic charge transfer salts based on the donor molecule bis(ethylenedithio)-tetrathiafulvalene, abbreviated as BEDT-TTF or ET, have been recognized as a highly correlated electron system.\cite{Kanoda1}  
Among them, $\kappa$-(ET)$_{2}$$X$ with $X =$ Cu(NCS)$_{2}$, Cu[N(CN)$_{2}$]$Y$ ($Y =$ Br and Cl), etc. has attracted considerable attention due to its strongly correlated quasi-two-dimensional electron system because the strong dimer structure consisting of two ET molecules makes the conduction band half filled.\cite{Kanoda2,Kino}  
Furthermore, the softness of the lattice in molecular conductors enables us to easily modulate the bandwidth by applying small physical and chemical pressures, while maintaining the band-filling.  
Therefore, the $\kappa$-(ET)$_{2}$$X$ system has been considered to be one of the typical bandwidth-controlled Mott systems.

The first-order Mott transition characterized by an $S$-shaped curve induces the correlated metal and superconductor phases from the Mott insulator phase by applying a small helium gas pressure or by the slight molecular substitution of the donor or anion molecules.\cite{Kanoda2,Ito,Lefebvre,Sasaki1,Limelette}  
The relation and transitions among the various phases in the phase diagram have been studied extensively, both experimentally and theoretically. 
Several unique features have been found in association with the Mott transition.
The first-order transition is terminated at $T_{\rm cr}^{\rm U} \simeq$ 32--40 K \cite{Lefebvre,Limelette,Kagawa1}, which is the upper critical end point and where novel criticality has been proposed on the basis of conductivity measurements.\cite{Kagawa1}
Recently, the intricate role of the lattice response at the Mott transition was emphasized and the different critical exponents were claimed.\cite{Souza} 
Along with the first-order phase transition, a macroscopic MI phase separation appears and induces an intrinsic electrical inhomogeneity.\cite{Sasaki2,Sasaki3}

One of the unsettled problems is the relation between superconductivity and the Mott insulator phase.  
Extensive studies have been carried out with various theoretical approaches,\cite{Murakami,Onoda,Watanabe,Powell} thereby increasing our understanding considerably.  
Most of them predict the anisotropic superconductivity to occur close to the Mott insulator phase. 
In contrast to the number of theoretical studies, the experimental investigations on this issue have not been carried out so much.  
Recently, a field-induced localization transition has been observed in the very narrow metallic region close to both the Mott insulating and superconducting phases.\cite{Taniguchi,Kagawa2} 
However, the role of the magnetic field in the correlation between the coherence of superconductivity and the Mott insulating order has not been clarified yet.  

In this paper, we present details of the $H$-$T$-bandwidth phase diagram for the organic Mott system $\kappa$-[($h$8-ET)$_{1-x}$($d$8-ET)$_{x}$]$_{2}$Cu[N(CN)$_{2}$]Br, where $h$8-ET and $d$8-ET denote fully hydrogenated and deuterated ET molecules, respectively. 
In the obtained phase diagram, the first-order Mott transition curve is found to be terminated at a lower critical end point in the upper critical field $H_{\rm c2}$-temperature plane of the superconductivity in addition to the previously observed upper critical end point.  

%Experiments

In order to tune the bandwidth by chemical pressures, single crystals of $\kappa$-[($h$8-ET)$_{1-x}$($d$8-ET)$_{x}$]$_{2}$Cu[N(CN)$_{2}$]Br that were partly substituted by deuterated ET molecules were grown by employing the standard electrochemical oxidation method.\cite{Yoneyama1}  
The substitution ratio $x$ denotes the nominal mole ratio to the fully deuterated ET molecule in the crystallization.\cite{substitution}
No apparent impurity effect in the electronic state other than the bandwidth modulation was observed for such a soft chemical substitution although a small increase in the scattering time within the clean-limit superconductivity was observed in the quantum magnetic oscillation experiments.\cite{Oizumi}

The in-plane resistivity in magnetic fields was measured by the dc four-terminal method, with the current parallel to the $a$-$c$ plane. 
The angle $\theta$ denotes the magnetic field direction relative to the $b$-axis, which is perpendicular to the $a$-$c$ plane.
The rotation axis was parallel to the current direction as shown in the inset of Fig. 2(b).
All the samples measured were cooled slowly at almost the same rate with $- 0.1$ K/min from 290 K to 80 K, $- 0.01$ K/min from 80 K to 70 K and keeping the temperature at 75 K for 900 min, and $- 0.1$ K/min from 70 K to 1.5 K.  
After the initial cooling, the temperature of the sample did not exceed 45 K during the resistivity measurements.

%Results and Discussion

\begin{figure}
\includegraphics[viewport=2cm 7cm 19cm 25cm,clip,width=0.8\linewidth]{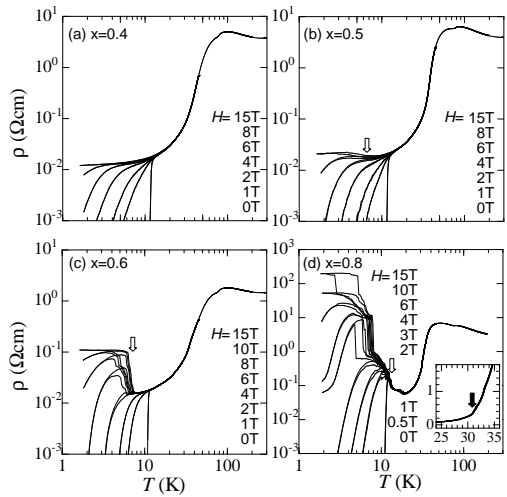}
\caption{Temperature dependence of the in-plane resistivity in magnetic fields perpendicular to the $a$-$c$ plane of $\kappa$-[($h$8-ET)$_{1-x}$($d$8-ET)$_{x}$]$_{2}$Cu[N(CN)$_{2}$]Br for (a) $x =$ 0.4, (b) $x =$ 0.5, (c) $x =$ 0.6, and (d) $x =$ 0.8.  White arrows indicate $T_{\rm MI}$. The inset of Fig. 1(d) shows $T_{\rm MI} \simeq$ 31 K (black arrow) at which the upper Mott MI transition occurs in addition to the lower $T_{\rm MI}$ transition due to the reentrance of the {\it S}-shaped curve.}
\end{figure}

Figure 1 shows the temperature dependence of the in-plane resistivity $\rho$ in magnetic fields perpendicular to the $a$-$c$ plane of $\kappa$-[($h$8-ET)$_{1-x}$($d$8-ET)$_{x}$]$_{2}$Cu[N(CN)$_{2}$]Br for (a) $x =$ 0.4, (b) $x =$ 0.5, (c) $x =$ 0.6 and (d) $x =$ 0.8.  
For the $x =$ 0.4 sample of Fig. 1(a), the $\rho$($T$) curve from room temperature to the superconducting transition temperature $T_{\rm c} \simeq$ 11 K represents the characteristic features observed in previous studies for the $x =$ 0 sample at $H =$ 0 T.
In addition, the usual normal state behavior appears at low temperatures; a constant residual resistivity of the metal is observed for $H > H_{\rm c2}$.  
In magnetic fields below $H_{\rm c2}$, the superconducting transition curves show typical fan-shaped broadenings, reflecting a vortex liquid state with large fluctuations.\cite{Lang,Kwok}
It may be difficult to evaluate the exact value of $T_{\rm c}$ and $H_{\rm c2}$ from the gradual transition curve of the resistivity in magnetic fields.  
This is basically because the change between the vortex liquid and normal states is not defined as a thermodynamic phase transition but observed as a crossover.\cite{Sasaki5}  
In this measurement, $T_{\rm c}$($H$) is defined as a temperature of initial decrease from the normal resistivity value.  
Although the $T_{\rm c}$ value determined by this means may have ambiguity in part, those are in good agreement with the mean field $T_{\rm c}$ obtained by the scaling analysis of fluctuations for the magnetization.\cite{Lang}

For the samples with $x \geq$ 0.5, resistivity anomalies at $T_{\rm MI}$ (marked with white arrows) are observed on the curves in the normal state at magnetic fields $H > H_{\rm c2}$ below $T_{\rm c}$($H =$ 0 T) in Figs. 1(b) ($x =$ 0.5) and 1(c) ($x =$ 0.6).  
In addition, a marked temperature hysteresis of the resistivity appears below $T_{\rm MI}$. 
With an increase in $x$, $T_{\rm MI}$ shifts to a higher temperature and the magnitude of the jump in the resistivity increases.  
In the case of the $x =$ 0.8 sample, $T_{\rm MI}$ already appears at $H =$ 0 T  above $T_{\rm c}$. 
Such anomalies associated with the jumps in and the hysteresis of the resistivity indicate that the first-order MI Mott transition occurs at $T_{\rm MI}$ in the normal states for $H > H_{\rm c2}$ ($x =$ 0.5 and 0.6) or $H =$ 0 ($x$ = 0.8).
The observed MI transitions are considered to be the same as those observed in the Mott insulator $\kappa$-(BEDT-TTF)$_{2}$Cu[N(CN)$_{2}$]Cl under hydrostatic helium gas pressure.\cite{Ito,Limelette,Kagawa2} 

In addition, another anomaly in the resistivity (marked with a black arrow in the inset of Fig. 1(d)) appears in the $x =$ 0.8 sample at around 30 K.  
The Mott transition curve of this system has been known to exhibit the characteristic $S$-shaped curvature.  
Hence, the observed anomaly at higher temperatures can be considered to result from the high $T$ paramagnetic insulator-metal (Fermi liquid)-low $T$ antiferromagnetic Mott insulator reentrant transition.  

A gradual decrease in the resistivity is found in the samples with $x$ ranging from 0.5--0.8 below $T_{\rm MI}$ for $H < H_{\rm c2}$, although the system should be in the Mott insulating state.  
This decrease may be due to the metal-insulator phase separation caused by the first-order Mott transition.\cite{Sasaki2,Sasaki3,Miyagawa}  
The metallic domains resulting from the phase separation become superconducting below $H_{\rm c2}$ and reduce the total resistivity of the dominant Mott insulator.
Therefore, on the resistivity curves for samples with $x$ ranging 0.5--0.8 under magnetic fields, the jumps in and the temperature hysteresis of the resistivity should be attributed to the Mott insulating domains, which are dominant in the bulk sample; moreover the decrease in the resistivity and its recovery by applying magnetic fields may be caused by the metallic/superconducting minor fragments. 

\begin{figure}
\includegraphics[viewport=2cm 5cm 18cm 25cm,clip,width=0.8\linewidth]{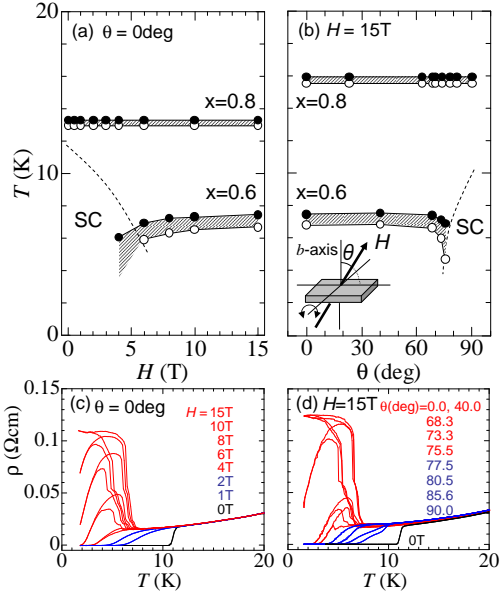}
\caption{(Color online) Metal-insulator transition temperature $T_{\rm MI}$ for the $x =$ 0.6 and 0.8 samples in magnetic fields: (a) $H$($\theta =$ 0 deg) and (b) $H$($\theta$) $=$ 15 T.  Open and closed circles denote $T_{\rm MI}$ determined from the down and up sweeps in temperature, respectively.  The shaded area indicates the temperature hysteresis of $T_{\rm MI}$.  The dashed curves represent $H_{\rm c2}$($T$) or $T_{\rm c}$($H$).  Inset of Fig. 2(b) shows the configuration of the sample and the magnetic field direction. Panels (c) and (d) denote the corresponding resistivity curves for the $x =$ 0.6 sample in $H$($\theta =$ 0 deg) and $H$($\theta$) $=$ 15 T, respectively.}
\end{figure}

The most important experimental finding of this study is that the jump in the resistivity disappears in a superconducting state with a finite resistivity due to the vortex flow, i. e., the vortex liquid state\cite{Kwok,Sasaki5,Sasaki4} below $H_{\rm c2}$, as shown in Figs. 1(b) and 1(c).
Figure 2 shows the magnetic-field dependence of $T_{\rm MI}$ and the dependence of $T_{\rm MI}$ on the magnetic field direction at low temperatures in the $x =$ 0.6 and 0.8 samples.\cite{T_MI_08}  
The open and closed circles in Figs. 2(a) and 2(b) denote $T_{\rm MI}$ determined from down and up sweeps of the temperature, respectively.  
The shaded area indicates the temperature hysteresis of $T_{\rm MI}$.
The dashed curve in Fig. 2(a) represents $H_{\rm c2}$($T$) or $T_{\rm c}$($H$) for the $x =$ 0 sample.\cite{Lang,Kwok}
In the $x =$ 0.8 sample, $T_{\rm MI}$, which is always higher than $T_{\rm c}$ does not change with the magnetic field and its direction.\cite{field_dep_T_MI}
In contrast, resistive anomalies such as the jump expected at $T_{\rm MI}$ for the $x =$ 0.6 sample are not observed on the smooth resistivity curve in the vortex liquid state below $H_{\rm c2}$ for $H =$ 2 and 1 T, as shown in Fig. 2(c).  
This clearly demonstrates that the MI transition does not occur in the vortex liquid state, i.e., $T_{\rm MI}$ starts to appear just above $H_{\rm c2}$($T = T_{\rm MI}$).  
It is noted that these phenomena require the condition $T_{\rm MI} \leq T_{\rm c}$ and the bulk superconductivity to be below $H_{\rm c2}$. 
A similar field-induced transition has been observed in pressurized $\kappa$-($h$8-ET)$_{2}$Cu[N(CN)$_{2}$]Cl.\cite{Sushko} 

In order to confirm the relationship between $T_{\rm MI}$ and $T_{\rm c}$ in $H$, the dependence of $T_{\rm MI}$ on the magnetic field direction was determined. 
The large anisotropy of $H_{\rm c2}$ for magnetic fields between the direction perpendicular and parallel to the $a$-$c$ plane enables us to control the $H_{\rm c2}$ value in a tilted magnetic field.  
The dashed curve in Fig. 2(b) shows $T_{\rm c}$ in magnetic fields based on the anisotropic effective mass model.\cite{Tinkham}
The resistivity curves in Fig. 2(d) indicate that $T_{\rm MI}$ appears only above $T_{\rm c}$ in the tilted constant magnetic field of 15 T.  
This is the same case of a varying magnetic field as shown in Fig. 2(c).
It can be concluded from both experiments on the dependence of $T_{\rm MI}$ on the magnetic field and its direction that the MI Mott transition does not occur in the superconducting state with a finite resistivity below $H_{\rm c2}$, i.e., the vortex liquid state.  

\begin{figure}
\includegraphics[viewport=2.5cm 6cm 17cm 24cm,clip,width=0.8\linewidth]{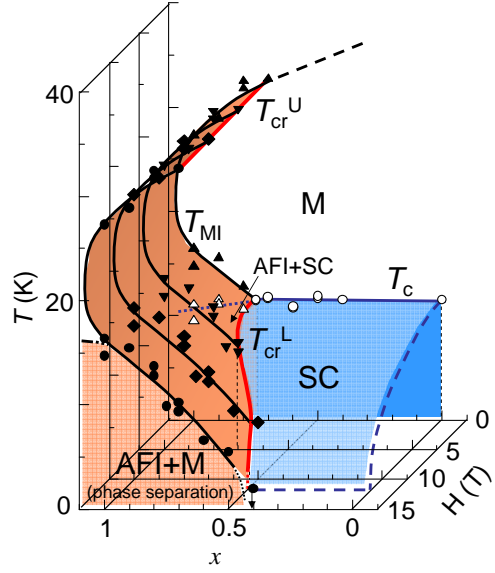}
\caption{(Color online) Temperature-magnetic field-substitution ratio $x$ phase diagram for $\kappa$-[($h$8-ET)$_{1-x}$($d$8-ET)$_{x}$]$_{2}$Cu[N(CN)$_{2}$]Br. The magnetic fields are applied perpendicular to the $a$-$c$ plane.  Filled and open symbols denote $T_{\rm MI}$ and $T_{\rm c}$, respectively. Thick red curves show $T_{\rm cr}^{\rm U}$ and $T_{\rm cr}^{\rm L}$.}
\end{figure}

Figure 3 shows the summary of $T_{\rm MI}$ and $T_{\rm c}$ in the $T$-$H$-$x$ phase diagram of $\kappa$-[($h$8-ET)$_{1-x}$($d$8-ET)$_{x}$]$_{2}$Cu[N(CN)$_{2}$]Br. 
The substitution ratio $x$ corresponds to the hydrostatic pressure $P$ (MPa) $= 35 - 15x$ for $\kappa$-($h$8-ET)$_{2}$Cu[N(CN)$_{2}$]Cl.\cite{Ito,Lefebvre,Limelette,Kagawa2}  
The value of $T_{\rm c}$ in magnetic fields is deduced from the magnetization and transport results reported so far\cite{Lang,Kwok}.
The overall features for $H =$ 0 T and above $H_{\rm c2}$ are similar to previous results for $\kappa$-($h$-ET)$_{2}$Cu[N(CN)$_{2}$]Cl under helium gas pressure by Kagawa {\it et al.}\cite{Kagawa2} and $\kappa$-($d$[$n, n$]-ET)$_{2}$Cu[N(CN)$_{2}$]Br by using partly deuterated ET molecules by Taniguchi {\it et al.}\cite{Taniguchi}
It must be emphasized that in the present phase diagram, the first order Mott transition curve is terminated also at the lower critical end point $T_{\rm cr}^{\rm L}$($H, x$) in the $H_{\rm c2}$($T$) plane in addition to the upper critical end point $T_{\rm cr}^{\rm U} \simeq$ 32--40 K.

Below $T_{\rm cr}^{\rm L}$ in the superconducting phase, the trace of the $S$-shaped curve of $T_{\rm MI}$($x$) is not observed, although it is difficult to isothermally investigate the transition property between the superconducting and the antiferromagnetic Mott insulating phases by means of the chemical substitution, as performed in the present study.  
Helium gas pressure experiments can approach this issue by varying the pressure isothermally.  
However, most of the previous studies have been performed for the MI transition above approximately 10 K for avoiding the solidification of helium, for example, the liquid-solid transition occurs at 7 K under 30 MPa.
Kagawa {\it et al.} have measured the isothermal pressure dependence of the resistivity at the MI transition in magnetic fields.\cite{Kagawa2} 
At the MI transition above $T_{\rm c}$($H$), the resistivity shows a clear sharp jump that represents the first-order transition.  
In addition, other previous results\cite{Taniguchi,Sushko} also have given some indications of the first-order transition at which a steep increase in the resistivity occurs in magnetic fields.  
On the other hand, below $T_{\rm c}$($H$) in the vortex liquid state, the behavior of the MI transition in the resistivity have become somewhat broad, which implies that the nature of the first-order transition becomes weak in spite of being at a lower temperature.\cite{Kagawa2}
The results reported so far may also be suggestive of the existence of $T_{\rm cr}^{\rm L}$.

\begin{figure}
\includegraphics[viewport=3cm 9cm 16cm 24cm,clip,width=0.8\linewidth]{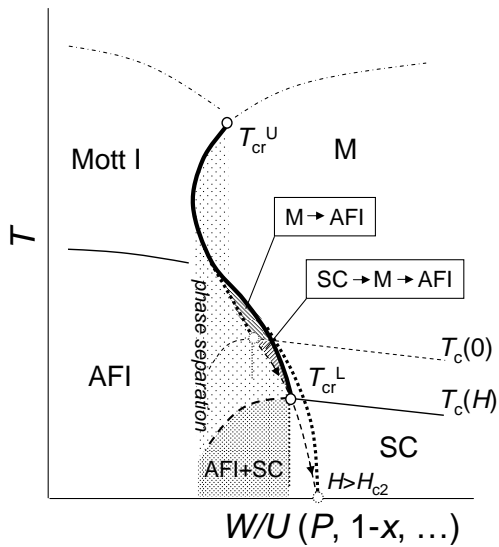}
\caption{Schematic of the effect of the magnetic field on the first-order Mott transition curve for $\kappa$-(BEDT-TTF)$_{2}$$X$.  The dotted area represents the macroscopic phase separation caused by crossing the first-order transition (thick line) during the temperature sweep.  For details, see the text.}
\end{figure}

Figure 4 schematically shows the effect of the magnetic field on the relationship between the Mott transition and superconductivity for magnetic fields $H < H_{\rm c2}$.  
On the basis of Fig. 4, we phenomenologically explain the behavior of the resistivity reported so far in a similar series of organic superconductors in the close vicinity of the Mott transition in magnetic fields.\cite{Taniguchi,Kagawa2}  
The lower critical end point $T_{\rm cr}^{\rm L}$($H, x$) of the first-order Mott transition curve moves to lower temperatures and toward a smaller $x$ (higher pressure) along with the suppression of $T_{\rm c}$ by magnetic fields.  
In $H > H_{\rm c2}$ in the end, the first-order transition curve may approach $T =$ 0. 
During this process of a decrease in $T_{\rm cr}^{\rm L}$ with an increase in the magnetic field, field-induced MI transitions occur in the hatched region shown in Fig. 4.  
In a part of the hatched region for $T_{\rm c}$($H$) ($= T_{\rm cr}^{\rm L}$) $< T < T_{\rm c}$(0), the successive superconducting-normal metal-insulator transitions occur due to the magnetic fields while only the MI transition appears above $T_{\rm c}$(0).   
Based on the present measurements, we would emphasize that there is no evidence for a direct transition from the superconducting to the insulating state by magnetic fields as it was claimed in Ref.\onlinecite{Taniguchi}

The existence of $T_{\rm cr}^{\rm L}$ in the $T_{\rm c}$($H$) plane may imply that the superconducting coherence developed below $T_{\rm c}$($H$) suppresses the Mott insulating phase at lower temperatures.  
Further theoretical investigations are necessary to understand the correlation between superconductivity and the Mott insulating states under magnetic fields.
Finally, we would like to comment on the reentrant AFI transition below $T_{\rm c}$, as claimed in the early helium gas pressure experiments on $\kappa$-($h$8-ET)$_{2}$Cu[N(CN)$_{2}$]Cl.\cite{Ito,Sushko,Posselt}  
The early experiments indicated the significant proximity of the AFI phase to the superconducting phase even for $H =$ 0.  
However, the reentrant transition always appeared around 7 K, even the different pressures in the range of 30--55 MPa.  
This difference may be caused by a drop or inhomogeneity in the actual pressure applied to the sample due to the solidification of liquid helium at this temperature (7 K)-pressure (30 MPa) range.

In summary, details of the phase diagram for the organic Mott system $\kappa$-[($h$8-ET)$_{1-x}$($d$8-ET)$_{x}$]$_{2}$Cu[N(CN)$_{2}$]Br in magnetic fields is presented.  
The lower critical end point of the first-order Mott transition, which is terminated in the $H_{\rm c2}$($T$) plane, is found experimentally in addition to the upper critical end point.  
A theoretical study on the stability of the Mott insulator and the adjacent superconductivity by considering the development of each order parameter and phase coherence in magnetic fields is necessary to understand this issue.

%acknowledgments
A part of this work was performed at HFLSM, IMR, Tohoku University.
This work was partly supported by a Grant-in-Aid for Scientific Research (Nos. 16076201, 17340099, and 18654056) from MEXT and JSPS, Japan.


\begin{thebibliography}{}
\bibitem{substitution} We checked the actual substitution with respect to the nominal value $x$ by measuring the vibrational mode of the terminal ethylene groups of the ET molecule in the infrared reflectance spectra.
\bibitem{T_MI_08} Somewhat different $T_{\rm MI}$ of the $x =$ 0.8 sample in Figs. 2(a) and 2(b) may be caused by the slight differences in the cooling conditions in the different series of the experiments.
\bibitem{field_dep_T_MI} Small change in $T_{\rm MI}$ could be seen in magnetic fields, which is related to the field-induced MI transition.\cite{Kagawa2,Taniguchi}  We, however, are not able to confirm the field dependece of $T_{\rm MI}$ in this study because of the low accuracy of the temperature measurements in magnetic fields in comparison to the expected tiny shift of $T_{\rm MI}$.
\end{thebibliography}

\end{document}